\documentclass[pra,superscriptaddress,twocolumn,showpacs]{revtex4}
\usepackage{epsfig}
\usepackage{amsbsy}
\usepackage{amsmath}
\usepackage{amsfonts}
\usepackage{graphics}
\usepackage{thumbpdf}
\usepackage{latexsym}
\usepackage{bm}

\begin{document}

\title{Chaos, entanglement and decoherence in the quantum kicked top}
\author{Shohini Ghose}
\affiliation{Dept. of Physics and Computer Science, Wilfrid Laurier
University, Waterloo, Canada}
\author{Rene Stock}
\affiliation{Dept. of Physics and Astronomy, University of Toronto, Canada}
\author{Poul Jessen}
\affiliation{College of Optical Sciences, University of Arizona, Tucson, USA}
\author{Roshan Lal}
\affiliation{Dept. of Physics and Computer Science, Wilfrid Laurier
University, Waterloo, Canada} 
\affiliation{Indian Institute of Technology, Kharagpur, India}
\author{Andrew Silberfarb}
\affiliation{Physical Measurement and Control, Edward L. Ginzton Laboratory, Stanford University, Palo Alto, USA}

\date{\today}
\begin{abstract}
We analyze the interplay of chaos, entanglement and decoherence in a system of qubits  whose collective behaviour is that of a quantum kicked top. The dynamical entanglement  between a single qubit and the rest can be calculated from the mean of the collective spin operators. This allows the possibility of efficiently measuring entanglement dynamics in an experimental setting. We consider a deeply quantum regime and show that signatures of chaos are present in the dynamical entanglement  for parameters accessible in an experiment that we propose using
cold atoms.  The evolution
of the entanglement depends on the support of the initial state on
regular versus chaotic Floquet eigenstates, whose phase-space distributions are concentrated on the corresponding regular or chaotic eigenstructures.  We include the effect of decoherence via a realistic model and show that the signatures of chaos in the entanglement dynamics persist in the presence of decoherence. In addition, the classical chaos affects the decoherence rate itself.
\end{abstract}

\maketitle

\section{Introduction}

In classical mechanics, the chaotic behavior predicted for non-integrable systems can be qualitatively different from the regular dynamics of integrable systems. The concept of regular versus chaotic dynamics at the quantum level has been more difficult to define as there is no clear measure of chaos in the quantum regime. Understanding the correspondence between quantum and classical evolutions in chaotic systems is a central problem in quantum mechanics and a major focus of the field of quantum chaos.  From a practical perspective, recent work~\cite{Shepelyansky} has shown that classical chaos can affect the implementation of quantum computing algorithms, and has fueled interest in identifying the effects  of chaos on quantum information theoretic properties such as entanglement and fidelity. Entanglement is thought to be a fundamental resource for many quantum information processing applications, and the effect of chaos on the dynamical generation of entanglement has been a topic of several studies~\cite{P08, entropy1, entropy2, entropy3, GS04, WGSH2003, L2003, J2003, chaos08}. The presence of chaos can also increase the rate of entanglement generation between a system and its environment~\cite{ZP95}, leading to increased decoherence and possibly stricter limitations on coherent quantum information processing.

In this paper, we explore the effect of chaos on entanglement and decoherence in a quantum kicked top.  We show that entanglement can be efficiently measured in this system, and identify signatures of chaos in the entanglement dynamics in a deeply quantum regime.  The quantum kicked top~\cite{QKT, QKT1}  has become a standard paradigm for theoretical studies of quantum chaos but has not yet been studied in experiments.  Here, we propose and analyze a possible experimental realization based on Cesium atomic spins  interacting with laser light and a pulsed magnetic field.  With a ground hyperfine spin of $F=4$ this system lies far from the semiclassical regime that is usually considered, since the size of $\hbar$ relative to the total phase space is roughly 0.1.  We study entanglement in the kicked top for parameters accessible with this system. Our analysis shows that signatures of chaos can be observed even if the system undergoes decoherence and that chaos affects the decoherence rate itself.  

A quantum kicked top with total angular momentum $j$ can be decomposed into  $N=2j$ spin-1/2 subsystems (qubits)~\cite{WGSH2003}.  We consider here the entanglement between a single qubit and the remaining $k=N-1$ qubits  (henceforth called $1:k$ entanglement). The  $1:k$ entanglement is of relevance to our proposed experiment with cold atoms because, as we show, it corresponds to entanglement between electron and nuclear spin in a single atom. Measurement of  entanglement  is usually experimentally challenging as it would require performing complete state tomography. For the case of the kicked top, we derive a simple expression for a $1:k$ entanglement measure in terms of the expectation values of the total spin of the $N$ qubits. Hence,  the $1:k$ entanglement can be efficiently monitored by simply
measuring the evolution of the mean total spin vector, thereby avoiding the need to perform complete state tomography. Although this result is generally applicable to any kicked top experimental realization, it is particularly useful for the cold atom experiments proposed here, in which the evolution of the mean spin vector can be efficiently measured in real time~\cite{Jessen1, Jessen2}.

We have performed numerical simulations of the kicked top dynamics to study the evolution of entanglement. Our results show that the long term $1:k$ entanglement dynamics exhibit signatures of classical chaos even in the extreme quantum regime considered here. Initial states localized in regular regions of the classical phase space exhibit quasiperiodic collapses and revivals, whereas initial localized states centered in chaotic regions quickly spread out and do not exhibit quasiperiodic entanglement dynamics.   These signatures are similar to those previously identified in the $2:k$ entanglement~\cite{WGSH2003} between 2 qubits and the rest in a semi-classical regime. We explain the entanglement behavior by extending the analysis in ~\cite{GS04} to the quantum regime. Quasiperiodicity versus irrregular dynamics depends on  the support of the initial state on regular versus chaotic Floquet eigenstates as shown in~\cite{GS04}. The differences in eigenstate decomposition for initial states centered in regular versus chaotic regions are limited by the small size of the spin considered here, and has no significant effect on the initial rate of entanglement generation as is the case for the semiclasscial regime~\cite{GS04}.  Nevertheless we find surprisingly clear signatures of regular versus chaotic dynamics in the long term evolutions. Regular and chaotic structures of the mixed classical phase space are also clearly evident
 in the Husimi distributions of the Floquet eigenstates in the quantum regime.
 
In addition to unitary evolution, we also consider the more realistic case of non-unitary dynamics when the kicked top system is coupled to its environment.
Our simulations of the entanglement dynamics include the effect of decoherence through a master equation description that accurately simulates photon scattering in our cold atom system.  Analysis of the negativity, a measure of entanglement for mixed states, shows that, although decoherence due to entanglement with the environment acts to reduce the overall $1:k$ entanglement, striking differences in the entanglement dynamics in regular versus chaotic regimes can persist for times longer than the decoherence time.  Furthermore, the rate of decoherence itself is slower in a regular regime than in a chaotic regime.  Signatures of chaos in the decoherence rate were previously analyzed in~\cite{ZP94, ZP95}  using a generic model of decoherence in a semiclassical regime.  Here we use a \emph{realistic} and accurate model to show that chaos affects the decoherence rate in a physical system in a deeply quantum regime.  Our proposed cold atom implementation would be the first to allow studies of decoherence and entanglement dynamics in a chaotic system.

The paper is organized as follows: In Section II we describe the basic features of the standard quantum kicked top and the corresponding classical system, which can exhibit chaos for certain dynamical parameters. Section III discusses the efficient measurement of $1:k$ entanglement dynamics and the possibility of experimental studies of entanglement with cold atoms. In Section IV we present an analysis of signatures of chaos in the entanglement dynamics of the
kicked top for a regime accessible in the cold atom system. We build on our previous analysis of the kicked top~\cite{GS04} and identify regular and chaotic eigenstates whose distributions are concentrated on the classical phase space structures in Section V. The relationship between decoherence, entanglement and chaos is analyzed in Section VI.
We present a summary and our conclusions in Section VII.

\section{The quantum kicked top and its classical limit}

The Hamiltonian for a quantum kicked top  is given
by~\cite{QKT,QKT1}
\begin{equation}
    H=\frac{\kappa}{2j\tau}J_x^2+pJ_y\sum_{n=-\infty }^{\infty }\delta
(t-n\tau ).  \label{top}
\end{equation}
Here, the operators $J_x$, $J_y$ and $J_z$ are angular momentum
operators obeying the commutation relation $[J_i,J_j]=\text{i}\hbar
\epsilon_{ijk} J_k$. The Hamiltonian describes a series of kicks
given by the linear $J_y$ term interspersed with torsions due to the
nonlinear $J_x^2$ term. The time between kicks is $\tau$, the angle
of turn per kick is given by $p$ and the strength of the twist is
determined by $\kappa$.  The magnitude $J^2=j(j+1)\hbar^2$ is a
constant of the motion

The classical map from kick to kick  can
be obtained from the Heisenberg evolution equations for the expectation values
of the angular momentum operators,
\begin{equation}
\langle J_i \rangle_{n+1} = \langle U^\dag J_{i} U \rangle_n.
\end{equation}
$U$ is the Floquet operator describing unitary evolution from kick
to kick,
\begin{equation}
U=\text{exp}(-\text{i}\kappa J_x^2/2j)\text{exp}(-\text{i}pJ_y),
\end{equation}
where the energy is henceforth rescaled such that $\tau=1$. 
The equations describing the evolution of the mean $J_x$, $J_y$ and $J_z$
involve functions of second moments of the angular momentum
operators, whose evolution in turn depends on third moments in an
infinite hierarchy of equations. In order to obtain the classical
mapping, we factorize all second and higher moments into products of
the mean values (1st moments) of the angular momentum operators.
This corresponds to localization of the state to a point in the
classical limit and breaks the infinite hierarchy of equations at
the level of first moments. Then, by defining the normalized
variables $X = \langle J_x \rangle /j$,  $Y = \langle J_y \rangle
/j$ and $Z = \langle J_z \rangle /j$, we can write down the
classical mapping
\begin{eqnarray}
X_{n+1} &=& X_n \text{cos} (p) + Z_n \text{sin}(p)\nonumber\\
Y_{n+1} &=&Y_n\text{cos}(\kappa \tilde{X}_n)-\tilde{Z}_n\text{sin}(\kappa\tilde{X}_n)\nonumber\\
Z_{n+1} &=& \tilde{Z}_n\text{cos}(\kappa \tilde{X}_n) + Y_n\text{sin}(\kappa\tilde{X}_n)\nonumber\\
\tilde{Z}_n&=&Z_n\text{cos}(p)-X_n\text{sin}(p) \nonumber \\
\tilde{X}_n&=&X_n\text{cos}(p)+Z_n\text{sin}(p).
\end{eqnarray}
$\tilde{X}_n$ and $\tilde{Z}_n$ are the angular momentum variables
subsequent to the kick but before the action of the twist.

The parameter $\kappa$ is the chaoticity parameter in the classical
kicked top. For $p=\pi/2$, Fig. 1 shows how $\kappa$ affects the kick-to-kick stroboscopic dynamics of the classical variables $\theta =
\text{cos}^{-1} Z, \phi = \text{tan}^{-1}(Y/X)$. When
$\kappa = 1$, the stroboscopic phase space is dominated by islands of regular
(periodic) motion [Fig.~1(a)]. As $\kappa$ is increased, the phase space
becomes mixed with regular islands embedded in a sea of chaotic trajectories [Fig~1(b)].  For even larger values of $\kappa$, the
system eventually becomes globally chaotic.

 \begin{figure}
\includegraphics[width=0.35\textwidth]{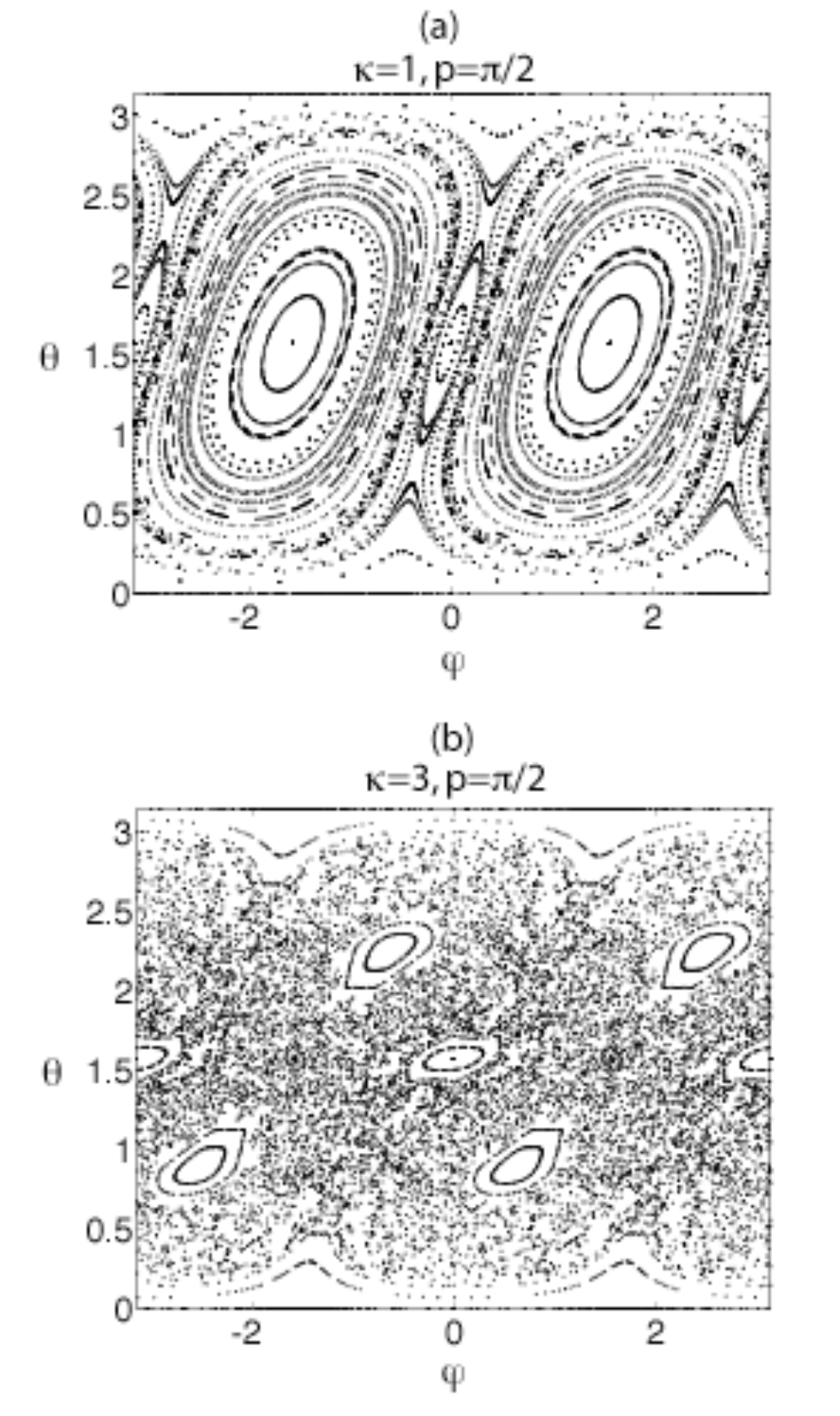}
\caption{Stroboscopic phase space maps at different values of $\kappa$ for kick strength $p=\pi/2$.  The classical spherical  coordinates $(\theta, \phi)$ are plotted after each kick  for 144 initial conditions, each evolved for 150 kicks. (a) For smaller $\kappa$, the phase space is dominated by regular orbits. (b)At larger $\kappa$, regular orbits are embedded in a sea of chaotic trajectories.}
\end{figure}

\section{Measurement of Entanglement in the kicked top}

\subsection{1:k entanglement measure}
We now focus on the dynamical evolution of
entanglement in the quantum kicked top. 
The total spin
$j$ can be considered as a system of $N=2j$ qubits with
\begin{equation}
 J_\alpha = \sum_{i=1}^{N} \frac{\sigma_{\alpha_i}}{2}, \,\, \alpha=x,y,z,
\end{equation}
where $\sigma_{\alpha_i}$ are the Pauli operators for the $i$th
qubit. For states in the symmetric subspace, the kicked top operator acts collectively on all N qubits
preserving the symmetry of the $N$ qubit state. This symmetry allows
us to write the spin expectation values for any single qubit as
\begin{equation}
\langle s_{\alpha}\rangle = \frac{\langle \sigma_{\alpha}
\rangle}{2} = \frac{\langle J_\alpha \rangle}{2j}.
\end{equation}

For an overall pure state, the entropy of the
reduced state $\tilde \rho$, of a single qubit is a measure of $1:k$ entanglement between a single qubit and the remaining $k=N-1$ qubits. For convenience we study the linear entropy,
$S = 1 - \text{Tr}[\tilde \rho^2]$, which ranges from 0 for
separable states to $1/2$ for maximally entangled states. The single qubit reduced state $\tilde \rho$ can in general be expressed as
\begin{equation}
\tilde \rho = \frac{1}{2} +  \langle \mathbf{s} \rangle \cdot \bm{\sigma},
\end{equation}
where $
\mathbf{s} $ is the mean spin vector of the qubit. Using Eq~(6), $\tilde \rho$ can be expressed in terms of the collective angular momentum operators as
\begin{equation}
\tilde \rho = \frac{1}{2} +  \frac{1}{2j}\langle \mathbf{J}\rangle \cdot \bm{\sigma}.
\end{equation}
The linear
entropy is then easily computed to be
\begin{equation}
S = \frac{1}{2}\left[1-\frac{1}{j^2}(\langle J_x\rangle^2+\langle
J_y\rangle^2+\langle J_z\rangle^2)\right].
\end{equation}
Measurement of the ensemble averaged angular momenta, $\langle J_\alpha
\rangle $, thus provides us with enough information to calculate $S$, which is a measure of the $1:k$ entanglement, without having to perform complete state tomography to reconstruct the entire multiqubit density operator. We also note that the function $S$ is identical to the generalized entanglement with respect to the angular momentum observables - an entanglement measure that is independent of subsystem division~\cite{YV06}. Furthermore, the linear entropy function is directly related to the `extent' or average spread of the state on the sphere. For a state that is highly localized on the sphere, the sum of the mean values of the angular momentum operators with be close to $j$  so the entropy $S$ will be close to 0. On the other hand, for a highly delocalized state, the mean values of the angular  momentum operators in Eq.~(9) approaches 0 and the value of $S$ approaches its maximum of $1/2$.

\subsection{Experiments with cold atoms}

An attractive physical system in which to realize a quantum kicked top is the total hyperfine (electron+nuclear) $\bm{F}$  of the electronic ground state of an individual atom.  In particular, we propose using samples of laser cooled alkali atoms to perform experiments on ensembles of identical kicked tops.  The atomic spins can be initialized by optical pumping, and manipulated in a controlled manner by Larmor precession in applied magnetic fields, and electric dipole interaction with an applied laser field.  Furthermore, one can perform polarization spectroscopy on the driving field to probe the spins in real time without perturbing the dynamics~\cite{Jessen1}.  These tools have already been used to demonstrate non-linear spin dynamics~\cite{Jessen2}, quantum state reconstruction~\cite{Jessen3} and quantum state control~\cite{Jessen4} with Cesium atomic spins. In the following, we briefly describe how to realize the kicked top with this system. 

In the low saturation, large detuning limit, the electric dipole interaction between a single atom and a monochromatic laser field is described by a light shift 
\begin{equation}\label{Dipole}
 U=-\frac{1}{4}\bm{E}^*\cdot \hat{\bm\alpha} \cdot \bm{E},
\end{equation}
where $\bm{E}=\text{Re}(\bm{E}e^{-\text{i}\omega t})$   is the electric field, and where the atomic polarizability  $\hat{\bm{{\alpha}}}$  is a rank-two tensor operator that can be decomposed into irreducible components of rank 0, 1 and 2~\cite{MOL, tensor}. Here, we consider alkali atoms restricted to a hyperfine ground state of given $F$.  The light shift $U$   is then an operator acting in a  $2F+1$-dimensional manifold, separable into three contributions from the irreducible components of   $\hat{\bm{{\alpha}}}$.  

The rank-0 contribution is a scalar interaction which does not couple to the spin degrees of freedom and therefore can be ignored.  The rank-1 contribution is an effective Zeeman interaction of the form $\bm{B}_{\text{eff}}\cdot \bm{F} $, where $\bm{B}_{\text{eff}}$  is proportional to the ellipticity of the laser field polarization.  For a linearly polarized driving field, this term disappears, leaving only the rank-2 contribution.  Choosing linear polarization in the x-direction the overall light shift is then quadratic in a component of the hyperfine spin as required for the kicked top, 
\begin{equation}\label{Dipole}
U=  \beta \hbar \gamma_s F_x^2.
\end{equation}
Here,  $\gamma_s=s\Gamma/2$ is the single-atom photon scattering rate, with the saturation parameter   $s$ depending on the laser intensity  $I$, detuning $\Delta$  , transition linewidth $\Gamma$   and saturation intensity  $I_0$ as $s=(\Gamma/2\Delta)^2(I/I_0)$.  The parameter $\beta$   is a measure of the relative timescales for unitary evolution and decoherence, and depends on the atomic species and the frequency of the driving field.  It takes on a maximum value of 8.2  for Cesium atoms driven at a frequency in-between the two hyperfine components of the D1 line at 894 nm~\cite{Jessen3}.  Comparing Eqs~(1) and(11), we see that the strength of the chaoticity parameter is related to the system parameters by
\begin{equation}
\kappa =2F\tau\beta\hbar\gamma_s.
\end{equation}
Thus larger values of  $\kappa$  are accompanied by higher rates of decoherence through photon scattering, and this limits the time over which one can observe unitary evolution of the system.  

The kicking term in the Hamiltonian of Eq. (1) can be implemented by a train of magnetic field pulses separated by a time $\tau$.  This results in a Zeeman interaction of the form  $g\mu_B\bm{B}\cdot \bm{F}\sum_n{\delta
(t-n\tau )}$ .  The finite bandwidth limitations of magnetic coils and drivers prevent the application of true  $\delta$-kicks, but it is not difficult in practice to keep the kick duration $T$  much shorter than the time  $\tau$ between kicks so that the  $\delta$-kick approximation remains valid.  The  angle of the turn per kick depends on the Larmor frequency of the applied magnetic field, 
\begin{equation}
p=\Omega_L T
\end{equation}
Hence, by adjusting the laser intensity, detuning, Larmor frequency, kick spacing and duration, one can explore a whole range of kicked top parameters  $\kappa$  and $p$.

To observe quantum dynamics in different regions of the classical phase space, we start with initial spin coherent states~\cite{SCSS}, which are rotations of the state $| j, m=j\rangle$ having maximum projection along the z-axis,
\begin{equation}
| \theta, \phi \rangle = \text{e}^{\text{i}\theta \left[J_x
\text{sin}\phi - J_y \text{cos}\phi \right]} |j, m=j \rangle.
\end{equation}
The expectation value of the spin in this state is given by $\langle \bm{J} \rangle = (j\text{sin}\theta \text{cos} \phi,  j\text{sin}\theta \text{sin} \phi, j\text{cos}\theta)$. In the qubit description, these states are separable with zero entanglement between qubits.  In our atomic system, the total angular momentum   $\bm{F}$ is the sum of a large nuclear spin ($I=7/2$ for Cesium) and a valence electron spin-1/2 system (qubit) whose reduced state can be described as in Eq. (7).  Starting with initial spin coherent states and restricting ourselves to the manifold of maximum $F$  ($F=4$  for Cesium), the quantity $S$   in Eq. (9) is then a measure of the entanglement between electron and nuclear spin, which can be experimentally measured by monitoring the mean collective angular momentum $\langle \bm{F} \rangle$  through, e. g., Faraday spectroscopy~\cite{Jessen1}.  Alternatively one can perform complete quantum state reconstruction of the overall electron+nuclear spin state~\cite{Jessen3}. This opens up the possibility of computing other entanglement measures such as negativity, and of monitoring decoherence by calculating the overall state purity.

\section{Dynamical Entanglement and chaos}
We analyze here
the $1:k$ entanglement dynamics of the kicked top described by the
Hamiltonian in Eq.~(1). We pick
the magnitude of the angular momentum $j$ to be 4 to correspond to
the $F=4$ hyperfine manifold in the Cesium ground state.
This puts the system far from the semi-classical regime of large $j$
$(j>100)$ that has been studied in previous work on
signatures of chaos~\cite{WGSH2003, GS04}. Even in
this deeply quantum regime with $j=4$, we can clearly identify the
effect of chaos in the entanglement dynamics.

For $\kappa
= 3, p=\pi/2$, we pick an initial spin coherent  state centered on a regular island of
Fig.~1(b) with $\theta =2.25 , \phi=2.5$. The resulting evolution of
the entanglement measure $S$ shows quasi-periodic behaviour [Fig.~2(a)] with collapses and revivals in the entanglement. In contrast,
for an initial state centered in the chaotic sea of  Fig.~1(b) with
$\theta =2.25 , \phi=1.1$, the quasiperiodic behaviour disappears
[Fig.~2(b)] and the evolution is irregular. 
For this same
initial state with $\theta =2.25 , \phi=1.1$, if we change $\kappa$
to 1, the classical dynamics becomes regular  again [Fig.~1(a)], and
correspondingly, quasiperiodic motion is recovered in the
entanglement dynamics in Fig.~2(c). For both the regular and chaotic dynamics, entanglement  at first increases as the initially localized state starts spreading over the phase space. However, for the states initialized in regular regimes, the dynamics causes the state to relocalize periodically, causing a reduction in entanglement. We explore this periodicity in more detail in the following section.

\begin{figure}
\includegraphics[width=0.475\textwidth]{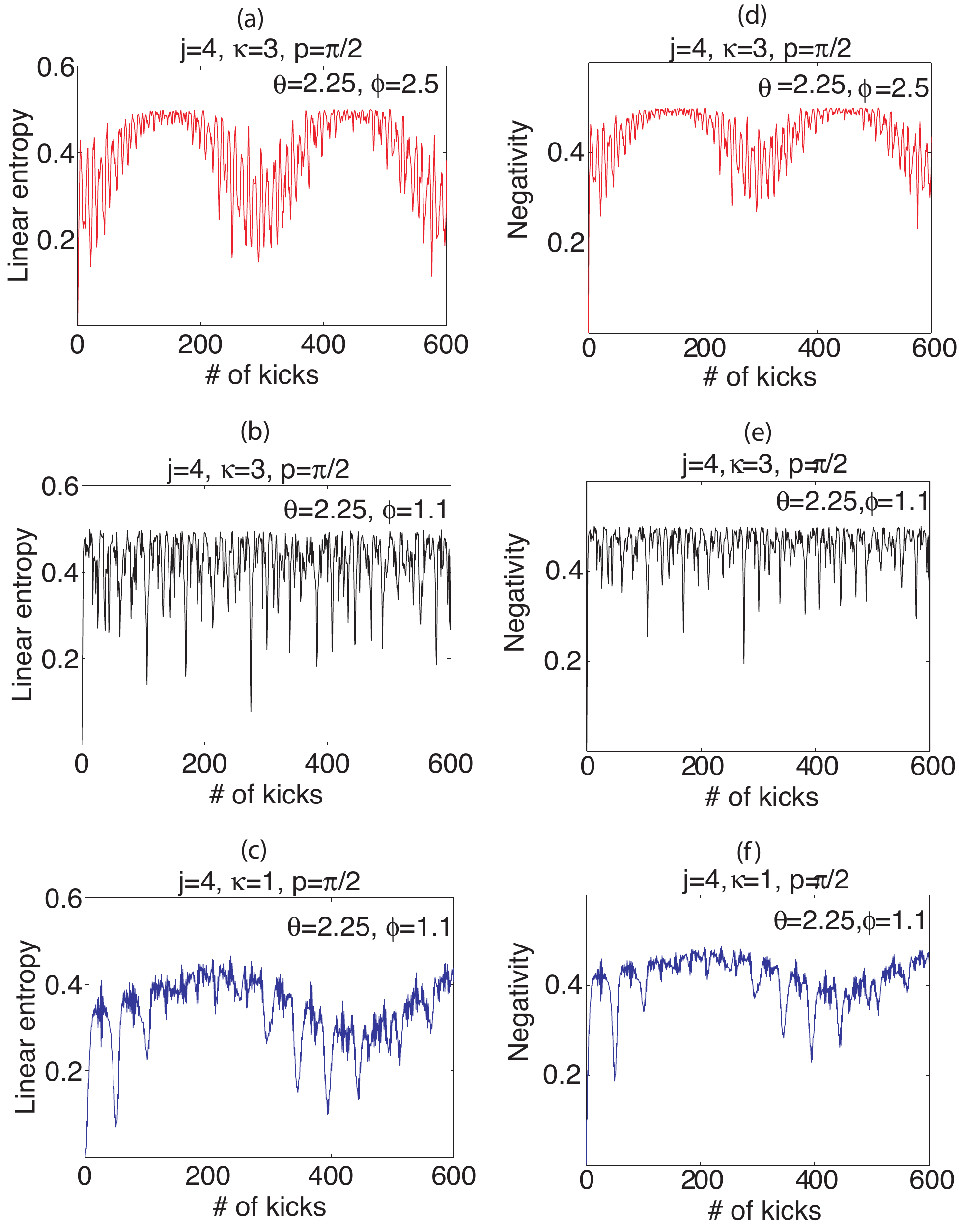}
\caption{Evolution of the linear entropy $S$, (a) - (c) and the negativity $N$, (d) - (f) for initial spin coherent states
$|\theta, \phi \rangle$ centered in regular versus chaotic regimes
of the classical phase space. See text for details.}
\end{figure}

We also compute a different measure of entanglement - namely the
negativity defined as~\cite{neg}
\begin{equation}
N = \frac{||\rho_T||-1}{2}
\end{equation}
where $\rho_T$ is the partial transpose of the overall system state
$\rho$, and the trace norm is defined to be
\begin{equation}
||\rho_T||=\text{Tr}\left[\sqrt{\rho_T^\dag \rho_T}\right].
\end{equation}
Entangled states
have a non-zero negativity, but the converse is not necessarily true
- i.e., states with zero negativity may nevertheless be entangled. Whereas entropy is a measure of entanglement only for pure
states, negativity can be used as an entanglement measure for mixed
states. We can thus use this measure later in Section V to understand the entanglement dynamics
when we include photon scattering in our model and the initial pure state becomes mixed due to decoherence. The evolution
of the negativity for the three initial states considered in Fig.~2(a)-(c) is shown in Fig.~2(d)-(f). The qualitative behavior exactly matches
the evolution of the  linear entropy. We can identify
collapses and revivals in the dynamics for initial states in a
regular regime and irregular evolution for states
initially in the chaotic sea. 

The three initial states analyzed above are representative of the general behavior of entanglement in regular versus chaotic regions. The entanglement dynamics changes more or less smoothly from regular quasiperiodic evolution to irregular evolution as we scan through initial spin coherent states from regular to chaotic regions. To illustrate this, Fig.~3(a) shows the average entanglement over $600$ kicks as a function of initial conditions $(\theta, \phi)$ scanned along the line $\theta=2.25$. The quasiperiodic behavior for initial states in the regular islands leads to lower average entanglement as compared to initial states in the chaotic sea. In Fig. 3(a), the regular islands can be clearly identified by the dips in the average entanglement. Conversely, in order to confirm that the entanglement behaviour is connected to the level of chaoticity of the classical map, we scan through the chaoticity parameter $\kappa$ for a fixed initial condition $\theta=2.25, \phi=1.1$. The time average entanglement as a function of $\kappa$ is shown in Fig.~3(b). As the chaoticity parameter is increased, the average entanglement increases. The dips in average entanglement reflect the occasional appearance of periodic orbits as fixed points in the classical phase space bifurcate in the approach to global chaos. 

\begin{figure}
\includegraphics[width=0.35 \textwidth]{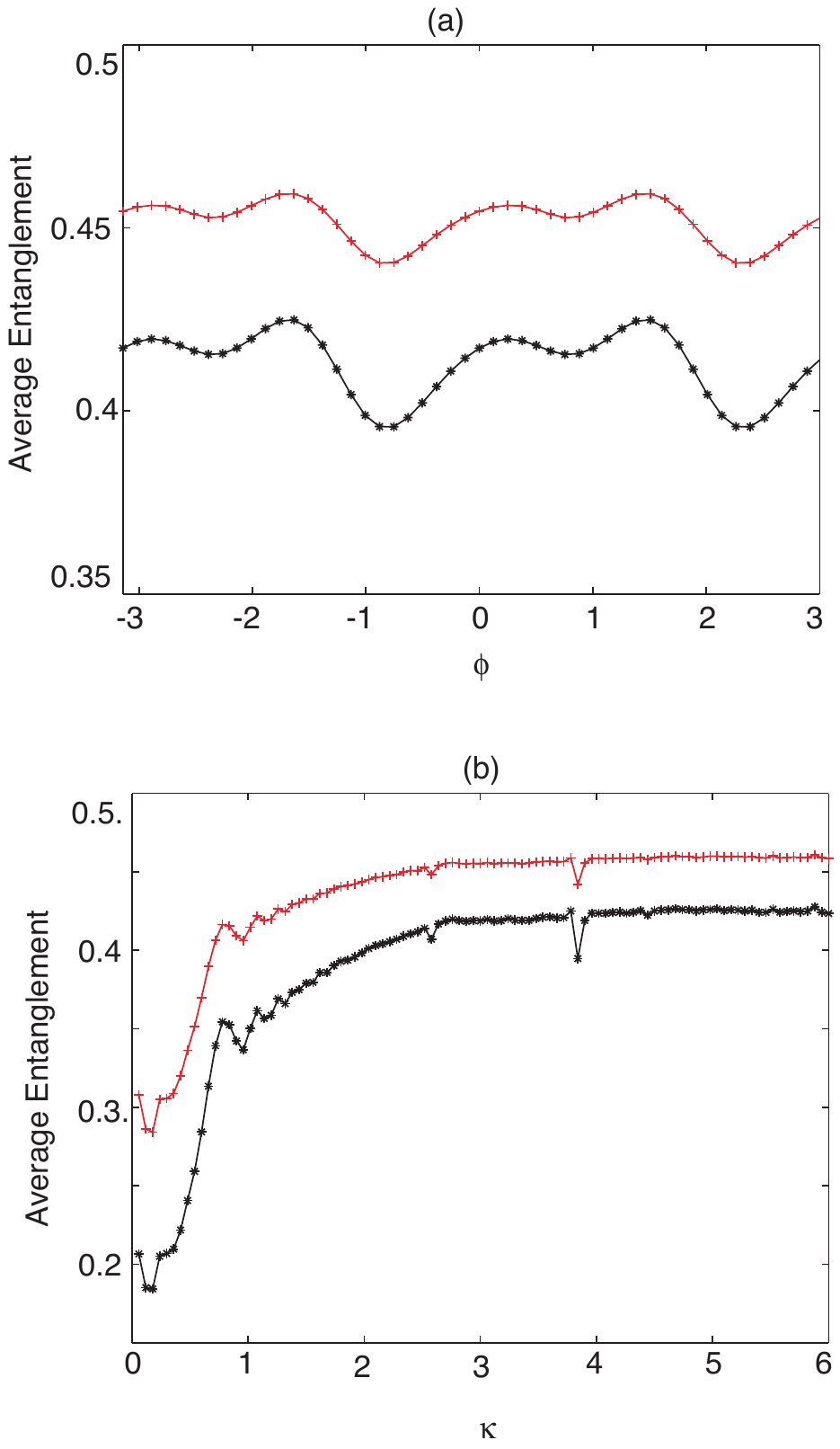}
\caption{Time averaged entanglement measured by both  entropy (stars) and negativity (plusses) reflects regular and chaotic classical structures as (a) the location of the initial spin coherent state is varied along the line $\theta=2.25, 0<\phi<2\pi$ for constant $\kappa=3, p=\pi/2$ and (b) $\kappa$ is varied keeping the initial state constant at $|\theta=2.25, \phi=1.1\rangle$. The average is taken over 600 kicks. For further details, see text.}
\end{figure}

\section{Analysis of Floquet Eigenstates} 

In previous work~\cite{WGSH2003}, we had identified similar
signatures of chaos in the $2:k$ entanglement dynamics of any two qubits
with the remaining qubits for a collection of $N > 50$ qubits. Here, we have
shown that these features are generic and can also be observed in
the $1:k$ entanglement between a single qubit and the rest. Furthermore,
these signatures are surprisingly persistent even for the case $j=4$ corresponding to only 8 qubits, which is far from the
semi-classical regime. Ref.~\cite{GS04} explained these universal
signatures of chaos by showing that in a semi-classical regime, the
eigenstates $|u_n\rangle$, of the kick-to-kick Floquet evolution operator (Eq.~3) can be classified as  `regular' or `chaotic'.
An initial state centered in a regular island has support on a few
almost degenerate regular eigenstates that give rise to
quasiperiodic motion involving a few regularly spaced eigenfrequencies, whereas a
state initially localized in a chaotic sea can be decomposed into a
number of chaotic eigenstates with a broad spectrum of incommensurate frequencies
contributing to the dynamics. 

Here, we test the above argument in the quantum regime. Figure 4 shows the eigenstate
decomposition
$f=|\langle u_n | \theta, \phi \rangle |^2$
 of the three initial states whose dynamics are shown
in Fig.~2. Each eigenstate is labelled by the corresponding eigenphase $\omega_n$,
\begin{equation}
U|u_n\rangle = \text{e}^{\text{i}\omega_n}|u_n\rangle,
\end{equation}
where $U$ is the Floquet evolution operator (Eq. 3).
The regular initial state of Fig.~2 (a) and (d) has
support on a few regular almost degenerate eigenstates [Fig.~4(a)], while the
chaotic initial state of Fig.~2 (b) and (e) has support on eigenstates with a broader
spectrum of irregularly space frequencies [Fig.~4(b)]. For the regular motion, the finite number
of frequencies present in the power spectrum of $S(t)$ can be
identified as sums of differences between eigenfrequencies of the regular
eigenstates on which the initial state has support~\cite{GS04}. In
the chaotic case, the large number of distinct eigenfrequency-difference sums
give
rise to a broader power spectrum. The regular state in Fig.~4(c) also shows support mostly on a single eigenstate and some support on almost degenerate eigenstate pairs. Although the spectrum looks similar to the chaotic state of Fig.~4(b), the larger number and irregular spread of eigenfrequencies in the chaotic state is sufficient to make the evolution much more irregular than the regular case. Upon closer examination, we find that unlike the chaotic case, the eigenfrequency spacing of  the dominant states in Fig.~4(c) is quite regular.  This regular spacing and the larger support on a single eigenstate leads to regular entanglement dynamics, with periodic revivals or `rephasing' of the evolutions of the different component eigenfrequencies, as seen in Fig.~2. Furthermore, support on the almost degenerate eigenstate pairs lead to long term periodic behaviour, which is missing in the chaotic case. The irregular oscillations on a fast timescale are due to the small but non-zero support on the remaining  eigenstates.

Due to the small value of
$j$ of 4, there are only a small total number of regular and chaotic
eigenstates, so the chaotic power spectrum will not be completely
flat and the dynamics will show some quasi-periodic behaviour. Furthermore, due to the mixed nature of the phase space, some eigenstates have overlap on regular as well as chaotic regions and can partially  contribute to initial states in both regular and chaotic regions. Thus the differences between regular and chaotic regimes are not as clearly delineated in this deeply quantum regime, compared to a semiclassical regime.
Nevertheless, the  regular and chaotic regions of the classical phase space can be  identified by the support of the initial  state on the Floquet eigenstates.  To verify this, in Fig.~5 we plot the quantity 
\begin{equation}
s=\sum_n{|\langle u_n | \theta, \phi \rangle |}
\end{equation}
as a function of initial conditions $(\theta, \phi)$ along the line $\theta=2.25$, for $\kappa=3, p=\pi/2$. This quantity roughly measures the number of eigenstates on which the initial state has support.  The dips in $s$ identify the regular regions in the mixed classical phase space and correlate perfectly with the time averaged entanglement.

\begin{figure}
\includegraphics[width=0.30\textwidth]{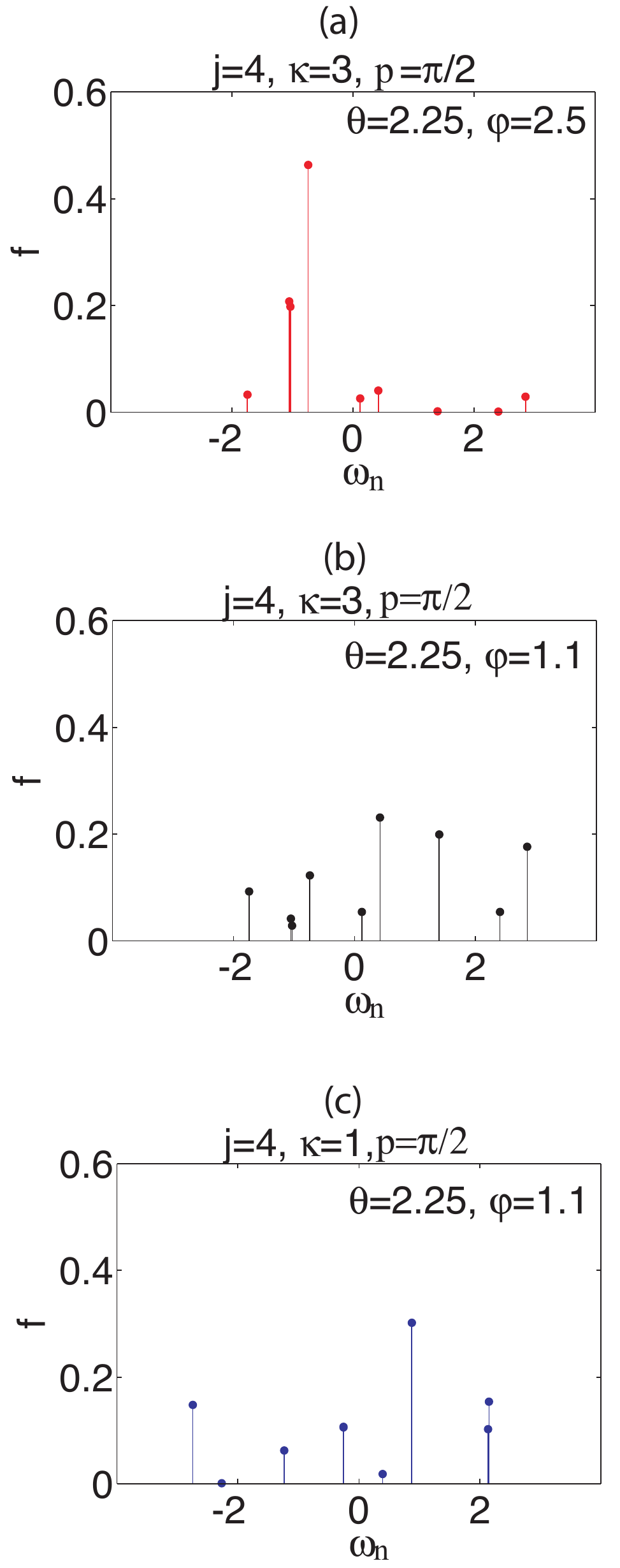}
\caption{Overlap $f=|\langle u_n | \theta, \phi \rangle |^2$ of
each of the 3 initial spin coherent states $|\theta, \phi \rangle$ in Fig. 2 with Floquet
eigenstates $|u_n\rangle$ corresponding to eigenfrequencies
$\omega_n$. Initial states in a regular region (a), (c) have support
on a few regularly spaced or degenerate eigenstates, while the state in the
chaotic sea (b), has support on a larger number of chaotic
eigenstates with irregularly spaced eigenfrequencies.}
\end{figure}

\begin{figure}
\includegraphics[width=0.30\textwidth]{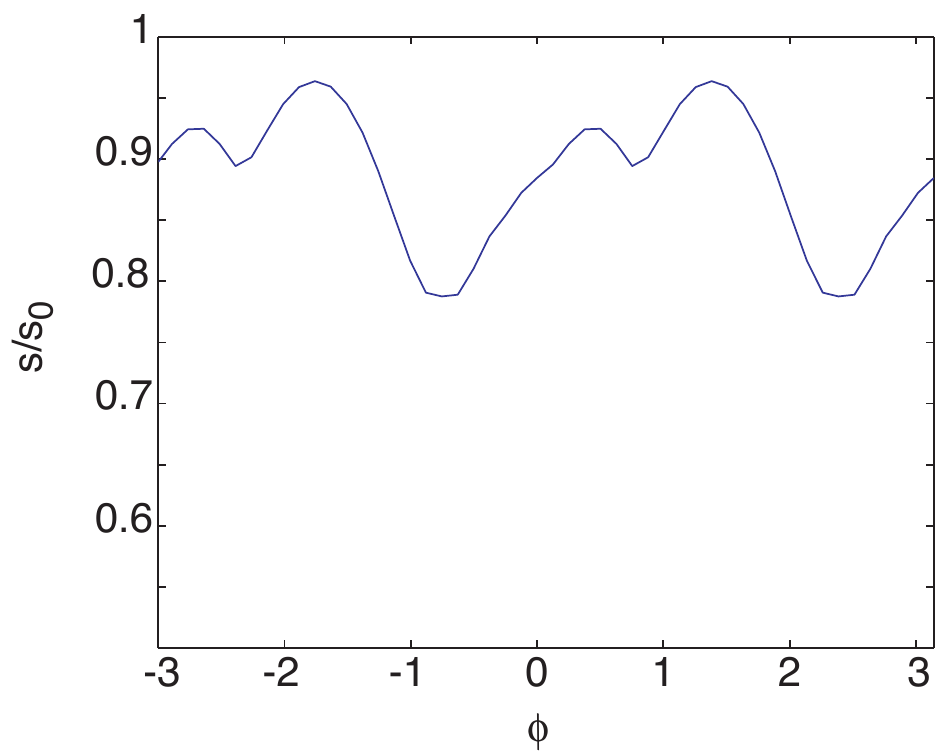}
\caption{Support  on the Floquet eigenstates, $|u_n \rangle$, $s=\sum_n{|\langle u_n | \theta, \phi \rangle |}$.  The initial spin coherent state $|\theta, \phi \rangle$ is varied along the line $\theta=2.25, 0<\phi<2\pi$ with fixed $\kappa=3, p=\pi/2$. $s$ is normalized with respect to its value of $s_0=3$, corresponding to the case of equal support over all eigenstates. The dips correspond to the location of the regular islands in the classical phase space of Fig.~1. }
\end{figure}

To further confirm the correspondence between the classical phase
space and the quantum eigenstates, Figure 6 shows the Husimi
quasiprobability distribution of the eigenstats with largest overlap with each of
the three initial states considered in Fig~2. The Husimi
distributions show overlap of a state  $|u_n\rangle$ with spin
coherent states,
\begin{equation}
P(\theta,\phi)=\frac{2j+1}{4\pi}\left| \langle \theta,\phi|u_n\rangle \right | ^2.
\end{equation}
The distributions clearly show that the
probability is concentrated on the classical phase space
structures shown in Fig~1. For regular eigenstates they are
concentrated along periodic orbits whereas for chaotic
eigenstates they are delocalized in the chaotic sea. Thus, although
quantum chaos typically deals with semi-classical techniques
to explore quantum-classical correspondence, our theoretical analysis provides evidence that very clear
signatures of classical chaos can be  present in regimes far from the
classical or even semi-classical regimes.

A second signature of classical chaos predicted by the semiclassical analysis is that the initial increase in entanglement is faster for the state starting in the chaotic sea than for those starting in regular regions~\cite{GS04}. In the chaotic regime the initial rate of increase was predicted to be exponential due to the contribution of a large number of incommensurate eigenfrequencies to the dynamics. In the quantum system considered here, due to the small size of the Hilbert space, there are not enough eigenfrequencies to give rise to an initial exponential increase in entanglement. Furthermore, the initial spin coherent states are not well localized in the phase space and therefore tend to have support on both regular and chaotic regions.  This in turn tends to wash out differences in the initial rate of entanglement generation. In fact, we find no significant difference in the initial increase in entanglement for states in regular versus chaotic regions.  Our studies thus not only serve to identify the correspondence between quantum and classical descriptions, but also highlight the differences inherent in quantum and classical systems.

\begin{figure}
\includegraphics[width=0.25\textwidth]{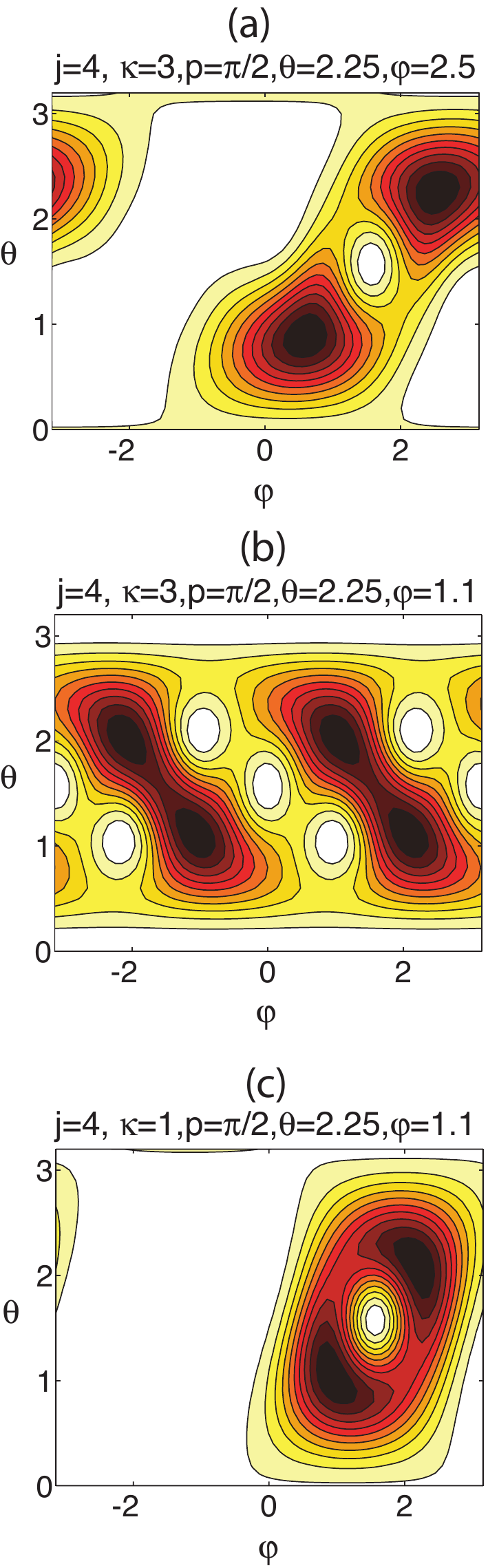}
\caption{Husimi distribution of the Floquet eigenstate having the
largest overlap with each of the three initial states (a)-(c) considered in
Fig.~2. Solid lines are contours of equiprobability distribution, and darker regions indicate larger probability.  Comparison with Fig.~1 clearly  shows concentration of the distributions along the classical
phase space structures.}
\end{figure}

\section{Dynamical Entanglement and decoherence}

In order to accurately model any physical system such as, e. g., Cesium atomic spins, 
we must include the effects of decoherence due to interaction with the environment. We perform simulations  for the specific case of the cold atom system including photon scattering which results in non-unitary evolution. We
model the evolution of the density operator $\rho$ of the system
using a master equation of the form
\begin{equation}
\frac{\text{d}\rho}{\text{d}t} =\frac{\text{i}}{\hbar}[H,\rho]
+\sum_q \left[ D_q\rho D^\dag_q-\frac{1}{2} \left\{ D^\dag_qD_q,\rho
\right \} \right].
\end{equation}
The jump operators $D_q=\epsilon_q\cdot \textbf{d}_{ge}$ describe
spontaneous emission from the excited states~\cite{Silberfarb}. The
excited states are adiabatically eliminated by assuming that the
ground state populations change slowly compared to the excited
states and the coherences. The excited state populations are
thus effectively slaved to the ground state populations to give a net
evolution only in terms of the ground state manifold $F=4$.
Furthermore, any population which is pumped into the $F=3$ ground
state is treated as loss and ignored in the subsequent dynamics.
This is possible if atoms in the $F=3$ manifold are not re-excited and therefore do not participate further in the observed dynamics. This model
has provided accurate simulations of the physical system in recent
experiments~\cite{Jessen2, Jessen3, Jessen4}. 

\subsection{Effect of decoherence on entanglement}
Figure 7 shows the simulated evolution (Eq.~20) of the entropy $S(t)$ including decoherence  for the three
initial states analyzed in Fig. 2. Comparing the unitary evolution of  Fig.~2(a)-(c) to the non-unitary evolution due to decoherence in 
Fig.~7 (a)-(c), we see that the qualitative features are the same. The
initial states in the regular regimes give rise to quasiperiodic
motion, which is missing in the chaotic regime. Thus the signatures
of chaos discussed in the previous section are also evident in the
open system dynamics. However, the total
time of evolution is limited by photon scattering, which causes the
system to get entangled with the environment. The linear entropy no longer quantifies the
$1:k$ entanglement of a single qubit (electron) with the rest (nucleus). Instead it
quantifies the total entanglement with both  the system and the
environment. As the system becomes more entangled with the
environment, there is an overall increase in entropy towards its maximum value of $0.5$.

\begin{figure}
\includegraphics[width=0.475\textwidth]{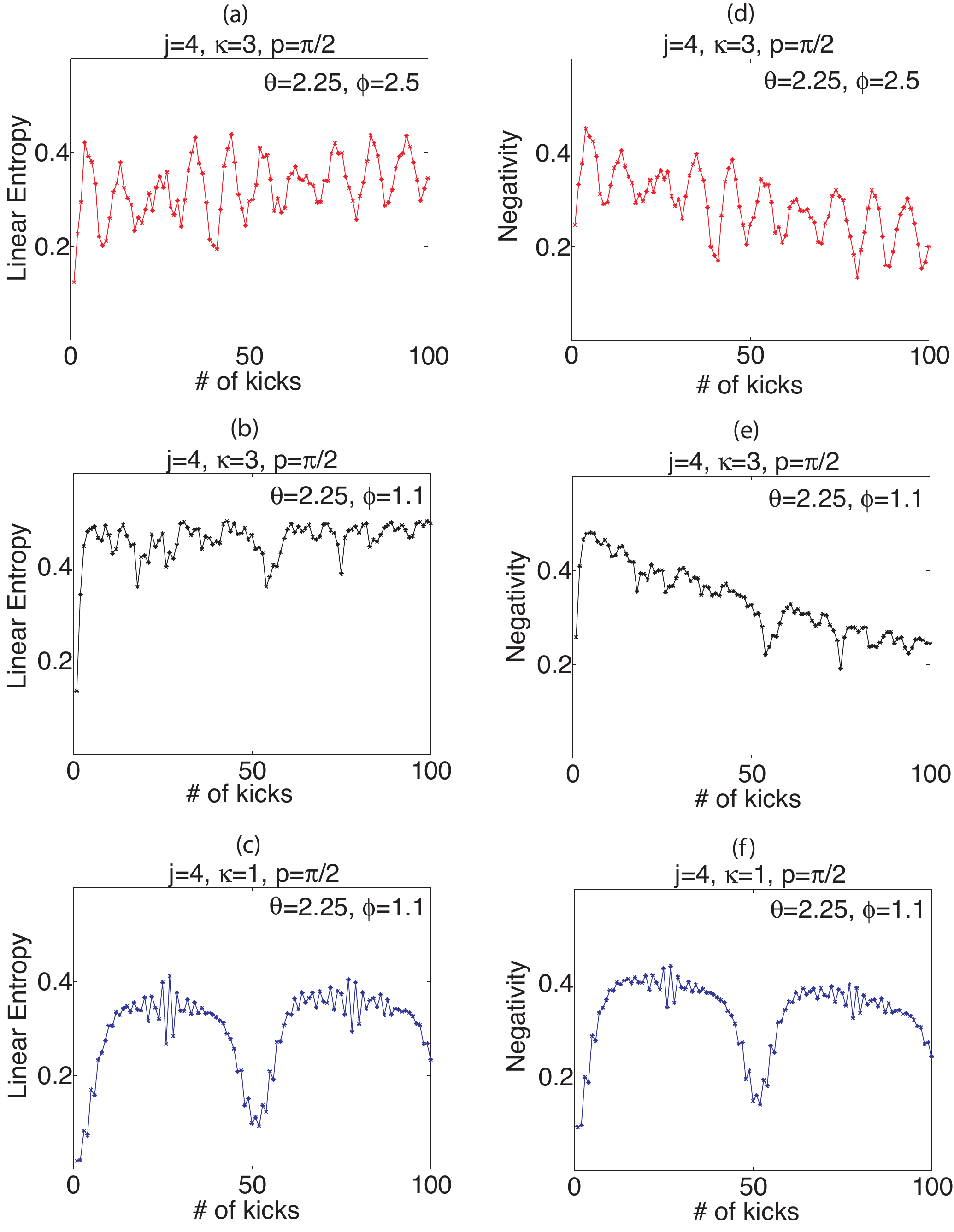}
\caption{Evolution of the linear entropy $S$, (a) - (c) and the negativity $N$, (d) - (f) in the presence of decoherence (photon scattering) for initial spin coherent states
$|\theta, \phi \rangle$ centered in regular versus chaotic regimes
of the classical phase space. For comparison, the 3 initial states considered are picked to be the same as those in Fig. 2. See text for details.}
\end{figure}

In order to better understand the effect of decoherence on the
signatures of chaos in the entanglement dynamics, we compute the
negativity as given by Eq.~(15). Unlike the entropy $S$, the negativity $N$ does give us a
valid measure of $1:k$ entanglement for an overall mixed state resulting from tracing out the
environment. Fig.~7(d)-(f) shows the evolution of the negativity for the
initial states considered in Fig. 2. We see that there is an
overall decrease in the  negativity but the quasiperiodic behavior
in the regular regimes and the irregular dynamics in the chaotic
regime exactly match the evolution of the entropy $S$ in Fig.~7(a)-(c).
This shows that the total spin state collectively
gets more entangled with the environment, and accordingly the total
$ 1:k$ entanglement within the system decreases. However, because decoherence acts collectively on the
\emph{total} spin system, the signatures of chaos caused by the
interactions \emph{within} the system are not completely destroyed
and persist well beyond the scattering time.

Although negativity has the advantage that it is a valid measure of entanglement for mixed states, it presents a disadvantage for experiments when compared to the linear entropy. This is because, in order to study negativity dynamics in an experiment, the total
(nuclear + electron) state must be tomographically reconstructed. This is experimentally more challenging than measuring the linear
entropy  via measurements of $\langle F_\alpha \rangle$ (Eq.~9). Hence, in actual experiments, it may be more convenient
only to measure the linear entropy $S$. Our comparison of the negativity
and the linear entropy (Fig. 7) then clarifies that the decoherence only
causes an overall decrease in the entanglement between electron and
nuclear spin. Thus the structures experimentally observable in the
linear entropy dynamics would correspond to valid signatures of
chaos in the entanglement between electron and nuclear spin. Furthermore the overall effect of the environment averages out in the time averaged entanglement dynamics, leading to curves very similar to the unitary evolution in Fig.~3.   

\subsection{Effect of chaos on decoherence rate}

In addition to examining the entanglement within the spin system, we can also analyze the entanglement of the total spin state with the environment due to photon scattering.
We set the
intensity, detuning and Larmor frequency to obtain $\kappa=3$ and
$p=\pi/2$. We then compare the decoherence rates  of a spin coherent
state centered on the regular island at $\theta=2.25, \phi=2.5$ and
a spin coherent state centered in the chaotic sea at $\theta=2.25$
and $\phi = 1.1$. Since the experimental parameters are the same for
both cases, the scattering rate $\gamma_s$  is also the same.
However, despite the scattering rate being constant, we find that
the decoherence rate is faster for the state initiated in the
chaotic sea relative to the state starting on the regular island. In
Fig. 8, we plot the purity of the overall state as a function of
time. The purity decay is a measure of the decoherence and shows
that the decoherence rate is faster for a state in the chaotic sea
than for a state on a regular island. We have observed this
increased rate of decoherence for states in a chaotic sea for other
values of $\kappa$ and $p$ as well, indicating that chaos generally
acts to increase the decoherence rate of the system.

\begin{figure}
\includegraphics[width=0.275\textwidth]{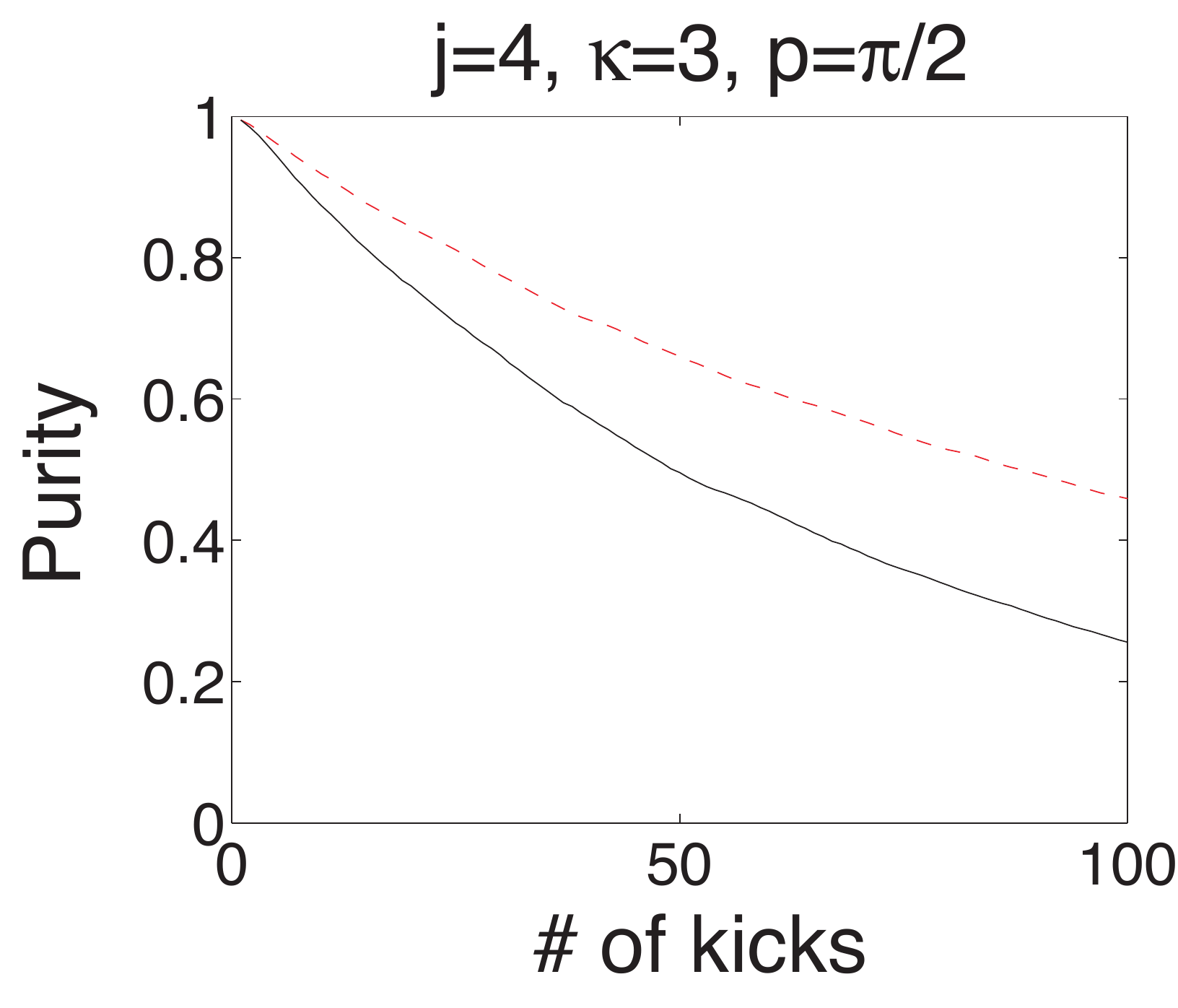}
\caption{Purity decay (decoherence) of the overall spin state for different initial conditions with fixed $\kappa=3, p=\pi/2$. The decoherence rate is faster for  
an initial spin
coherent state centered on a regular island with $|\theta=2.25, 
\phi=2.5 \rangle$ (dashed) compared to an initial spin coherent
state centered in the chaotic sea with $|\theta=2.25, \phi=1.1
\rangle$ (solid), although  the photon scattering rate is the same for both cases.}
\end{figure}

The increase in the decoherence rate in the chaotic regime can be
understood by examining the Lindblad terms in the master equation of
Eq.~(20) more carefully. The
populations in the excited states which determine the spontaneous
emission rate, depend on the ground state populations. Figure 9 shows
the Husimi distributions after $50$
kicks for the two initial spin coherent states centered in the
regular island and chaotic sea.  Whereas, the initial state on a regular island remains close to a more robust spin coherent state, for the initial state in the chaotic sea,  the dynamics causes the
population distribution to quickly lose any  symmetry and diverge from a coherent state. The spontaneous emission for such a state leads to a net larger decoherence rate.

\begin{figure}
\includegraphics[width=0.25\textwidth]{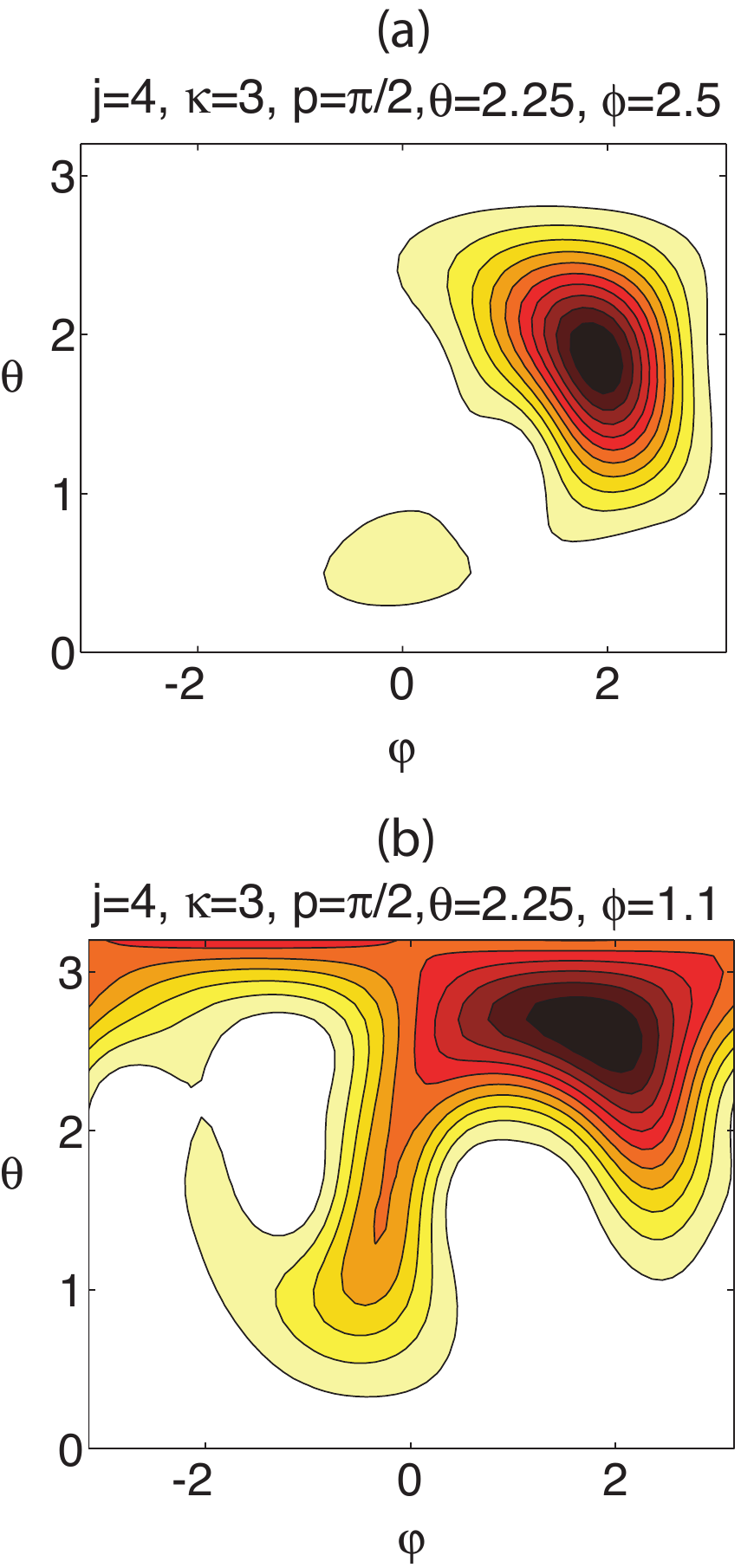}
\caption{Husimi distributions after 50 kicks for the initial states considered in Fig. 8, with $\kappa=3, p=\pi/2$. Solid lines are contours of equiprobability distribution, and darker regions indicate larger probability. The initial spin coherent state centered on a regular island (a) with
$|\theta=2.25. \phi=2.5 \rangle$ remains closer to a robust spin coherent state, whereas
the  initial spin
coherent state centered in the chaotic sea with $|\theta=2.25.
\phi=1.1 \rangle$ spreads quickly, leading to a delocalized state with net larger spontaneous emission and hence larger decoherence.}
\end{figure}

Our results support the arguments originally formulated by Zurek and Paz~\cite{ZP94, ZP95}
claiming that chaos can enhance the rate of decoherence. The relevance of our work is that we consider a realistic and accurate decoherence model for an actual physical system very deeply in the quantum regime and verify the predictions
obtained from semiclassical theories and general decoherence
models.

\section{Conclusion}

In summary, we have presented an analysis of entanglement and decoherence in the quantum kicked top in a regime of relevance to a proposed experiment with cold Cesium atoms. $1:k$ entanglement can be easily calculated from the mean  angular momentum vector, and signatures of chaos are
evident in the dynamics of entanglement in a quantum regime.  The dynamics  can
be understood by examining the decomposition of the initial state
into regular and chaotic eigenstates of the Floquet  operator.

Our simulations of a realistic system of cold atoms undergoing decoherence due to photon scattering  reveal that
decoherence reduces the $1:k$ entanglement but does not erase the signatures of chaos.
Furthermore, we have shown that chaos can enhance the overall
decoherence rate, or entanglement of the system with the
environment. This shows that whereas chaos can be helpful for
generating entanglement within the system, it can have negative
effects by rapidly causing decoherence.

We have thus shown a means for performing the first experimental studies of entanglement in a chaotic system, identified signatures of chaos in entanglement in a
quantum regime that previous theoretical studies have not
investigated, and extended the previous theoretical analysis to this
quantum regime. Furthermore, we have demonstrated the collective effect of
decoherence on the entanglement, and identified and explained
signatures of chaos in the decoherence rate using a realistic model
of decoherence.  An effort to implement the experiments decribed here is
currently underway.

\acknowledgements We thank I. Deutsch for discussions and insights. SG  was supported by an NSERC
Discovery Grant.

\end{document}